\documentclass[twocolumn,secnumarabic,amssymb,amsmath,superscriptaddress,aps,longbibliography]{revtex4-1}

\usepackage{hyperref}
\usepackage{graphicx}
\usepackage{amsfonts}
\usepackage{amssymb}
\usepackage{bm}
\usepackage{color}
\usepackage{xspace}
\usepackage{cancel}
\usepackage{braket}
\usepackage{chemformula}
\usepackage{array}
\usepackage{ifthen}
\newcommand*\E{\mathrm{e}}
\newcommand*\ii{\mathrm{i}}

\allowdisplaybreaks

\begin{document}

\title{Theory of nuclear motion in RABBITT spectra}

\author{Serguei Patchkovskii}
\email{serguei.patchkovskii@mbi-berlin.de}
\affiliation{Max-Born-Institute, Berlin, Germany} 
\author{Jakub Benda}
\affiliation{Institute of Theoretical Physics, Faculty of Mathematics and Physics, Charles University, Prague, Czech Republic}
\author{Dominik Ertel}
\affiliation{Albert-Ludwigs-Universität Freiburg, Germany}
\author{David Busto}
\affiliation{Albert-Ludwigs-Universität Freiburg, Germany}
\affiliation{Department of Physics, Lund University, P.O. Box 118, 22100 Lund, Sweden}


\date{\today}

\begin{abstract}
Reconstruction of attosecond beating by interference of two-photon transitions (RABBITT) is a
powerful photoelectron spectroscopy, offering direct access to internal
dynamics of the target. It is being increasingly applied to molecular systems,
but a general, computationally tractable theory of RABBITT spectra in molecules
has so far been lacking. We show that under quite general assumptions, RABBITT
spectra in molecules can be expressed as a convolution of the vibronic
cross-correlation functions and two-electron photoionization matrix elements.
We specialize the general expressions to the commonly-encountered special
cases. We expect our theory to enable accurate modeling and interpretation of
molecular RABBITT spectra in most medium-sized molecules.
\end{abstract}
\maketitle

\section{Introduction\label{sec:intro}}

Reconstruction of attosecond beating by interference of two-photon transitions (RABBITT) is an
ingenious photoelectron spectroscopy, offering direct access to the
photoelectron phases\cite{Paul2001}, and consequently to the intricate details
of the electronic and nuclear dynamics in atoms and
molecules\cite{Isinger2017,Vos2018,Cattaneo2018}. RABBITT spectroscopy is being
increasingly applied to molecular systems, with many notable recent theory
developments (see e.g.
\cite{Ahmadi2020,Ahmadi2022,Carpeggiani2017,Benda2022,Cattaneo2018} and
references therein). An essential factor, affecting all molecular
spectroscopies, is nuclear motion, which however received only limited
attention in the literature so far\cite{Cattaneo2018,Haessler2009,Nandi2020}.
This oversight is likely partially due to the enormous cost of the brute-force
treatment of the electron-nuclear coupling in photoionization, which so far
limited practical calculations to a very few nuclear degrees of freedom. A
similar difficulty arises in molecular spectroscopy of bound-to-bound
transitions, where it been long recognized\cite{Heller1981,Tannor2006} that the
problem can nonetheless be made numerically tractable, by recasting it in a
time-dependent form. The observable effects of the nuclear motion are then
compactly summarized by vibronic auto- and cross-correlation
functions\cite{Heller1981,Tannor2006}. The utility of the nuclear auto- and
cross-correlation functions have also been recognized in the strong-field and
atto-second domain, where they have been used to describe nuclear-motion
effects in high-harmonics generation\cite{Lein2005,Baker2006,Patchkovskii2009} and
attosecond electron-hole migration\cite{Vacher2017,Arnold2017,Ruberti2022}.
Very recently, an elegant theory of the molecular electron-streaking spectra
has been developed\cite{Kowalewski2016}, with the single-surface nuclear
autocorrelation functions taking the central role. 

In this contribution, we extend the approach of Ref.~\cite{Kowalewski2016} to
the theory of molecular RABBITT photoelectron spectra. We derive a compact,
general expression for the relevant transition amplitudes in terms of vibronic
cross-correlation functions. Our treatment includes, at least in principle, all
nuclear motion effects relevant for RABBITT transitions. In particular, it
describes the effects of the coherent averaging over the initial vibrational
function, including the zero-point effects; the redistribution of the absorbed-
photon energy between the photoelectron and internal degrees of freedom; the
effects of absorption and emission of additional IR photons by the cationic
core; the effects of the finite pulse duration. The treatment naturally
includes complex vibronic dynamics in the vicinity of conical intersections as
well.

The rest of this manuscript is organized as follows: The following section
\ref{sec:theory} develops the general theory of RABBITT spectra in molecules.
Section~\ref{sec:special-cases} considers some relevant special cases, which
allow further simplifications of the general expression. Finally,
section~\ref{sec:summary} summarizes the work, and presents an outlook for
follow-up investigations and applications.

\section{Theory\label{sec:theory}}

We are interested in modeling photoelectron spectra of a molecular system,
described by a field-free, time-independent Hamiltonian $\hat{H}_0$, produced
as a result of interacting with a three-colour laser field. The individual
components of the field, all taken to be linearly-polarized, are given by:
\begin{align}
  F_{IR}\left(t\right) &= f_{IR}\left(t\right) \vec{n}_{IR} \cos\left(\omega t\right) & \label{eqn:fIR} \\
  F_{i} \left(t\right) &= f_{i} \left(t\right) \vec{n}_{i} \cos\left(\Omega_i t + \Phi_i\right), & \label{eqn:fi}
\end{align}
where $i=1,2$, $\vec{n}$ is field polarization direction, and $f\left(t\right)$
is a slowly-varying envelope.  The corresponding terms in the total Hamiltonian
are given by:
\begin{align}
  \hat{V}_{IR} &= \frac{1}{2} \hat{\mu}_{IR}  f_{IR} 
                  \left(t\right) \left[ \E^{ \ii\omega t} 
                                      + \E^{-\ii\omega t} \right]  & \label{eqn:VIR} \\
  \hat{V}_{i}  &= \frac{1}{2} \hat{\mu}_{i} f_{i}\left(t\right) 
                  \left[ \xcancel{\E^{ \ii\Omega_i t + \ii\Phi_i}} 
                                + \E^{-\ii\Omega_i t - \ii\Phi_i} \right], & \label{eqn:Vi}
\end{align}
where the terms in brackets correspond to the emission ($+$) and absorption
($-$) of a photon. We assume that the field parameters are such that only
absorption is possible for the XUV fields ($\hat{V}_{i}$), while the IR photons
can be both absorbed and emitted. Operators $\hat{\mu}$ incorporate the specific
form of the field-interaction Hamiltonian and field polarization properties.
Eqs.~\eqref{eqn:VIR},\eqref{eqn:Vi} implicitly assume that the laser field is
treated in the length gauge and dipole approximation.

In addition to the field-free Hamiltonian $\hat{H}_0$, we will also consider
Hamiltonians $\hat{H}_{IR}$ and $\hat{H}_{i}$, defined as:
\begin{align}
  \hat{H}_{IR} &= \hat{H}_0 + \hat{V}_{IR},               & \label{eqn:HIR} \\
  \hat{H}_{i}  &= \hat{H}_0 + \hat{V}_{IR} + \hat{V}_{i}, & \label{eqn:Hi}
\end{align}
corresponding to our preferred order of treating the perturbations. For each
Hamiltonian $\hat{H}_a$, the corresponding propagator $\hat{U}_a$ is
symbolically given by:
\begin{align}
  \hat{U}_a\left(t',t;E\right) &= \E^{-\ii\int_t^{t'} dt" \left(\hat{H}_a-E\right)}, & \label{eqn:Ua}
\end{align}
where we have chosen to pull the rapidly-oscillating phase $\E^{-\ii E t}$ out.
The ``characteristic energy'' $E$ is in principle arbitrary; however, we expect
that it is selected such as to make $\hat{U}_a$ a slow function of time.  We
note that a propagator in Eq.~\eqref{eqn:Ua} satisfies the energy-origin
transformation:
\begin{align}
  \hat{U}_a\left(t',t;E\right) &= 
     \E^{\ii\left(E-E'\right)\left(t'-t\right)}
     \hat{U}_a\left(t',t;E'\right)
     . & \label{eqn:Ua-origin}
\end{align}

\subsection{Wavefunction response to a two-colour field}

Our first task is to calculate the wavefunction response to the combined
effects of the IR and \textit{one} of the XUV fields. We will only consider the
contribution bilinear in the two fields, and assume that the contributions due
to each field alone can be neglected (e.g. because they are energetically
separated).

We start by treating $\hat{H}_{IR}$ as the zeroth-order Hamiltonian, and
$\hat{V}_{i}$ as the perturbation. The usual time-dependent perturbation
theory then yields\cite{Tannor2006,Milosevic2006,Kulander1978}:
\begin{widetext}
\begin{align}
  \ket{\Psi_{i}^{(1)}\left(t\right)} 
     &= -\ii \int_{t_0}^{t} dt' 
         \E^{-\ii E_{I}\left(t-t'\right)}\hat{U}_{i}\left(t,t';E_{I}\right)
         \hat{V}_{i}\left(t'\right) 
         \E^{-\ii E_{N}\left(t'-t_0\right)}\hat{U}_{IR}\left(t',t_0;E_{N}\right) 
         \ket{\Psi^{(0)}\left(t_0\right)}
     & \nonumber \\
     &= -\ii \int_{t_0}^{t} dt' 
         \E^{-\ii E_{I}\left(t-t'\right)}\hat{U}_{i}\left(t,t';E_{I}\right)
         \frac{1}{2} \hat{\mu}_{i} f_{i}\left(t'\right) \E^{-\ii\Omega_i t' - \ii\Phi_i}
         \E^{-\ii E_{N}\left(t'-t_0\right)}\hat{U}_{IR}\left(t',t_0;E_{N}\right) 
         \ket{\Psi^{(0)}\left(t_0\right)}
     & \nonumber \\
     &= -\frac{\ii}{2} \E^{-\ii\Phi_i +\ii E_N t_0}
         \int_{t_0}^{t} dt' 
         \E^{-\ii E_I t}
         \E^{\ii \left(E_{I}-E_{N}-\Omega_{i}\right) t'}
         f_{i}\left(t'\right)
         \hat{U}_{i}\left(t,t';E_{I}\right)
         \hat{\mu}_{i} 
         \hat{U}_{IR}\left(t',t_0;E_{N}\right) 
         \ket{\Psi^{(0)}\left(t_0\right)}
     & \nonumber \\
     &\approx 
        -\frac{\ii}{2} \E^{-\ii\Phi_i +\ii E_N t_0}
         \int_{t_0}^{t} dt' 
         \E^{-\ii E_I t}
         \E^{\ii \left(E_{I}-E_{N}-\Omega_{i}\right) t'}
         f_{i}\left(t'\right)
         \hat{U}_{IR}\left(t,t';E_{I}\right)
         \hat{\mu}_{i} 
         \hat{U}_{IR}\left(t',t_0;E_{N}\right) 
         \ket{\Psi^{(0)}\left(t_0\right)}
     & \label{eqn:psi1t}
\end{align}
\end{widetext}
where $E_N$ and $E_I$ are respectively characteristic energies of the system
before and after XUV photon absorption, and $t_0$ is chosen before the start of
the XUV pulse [i.e. $f_i\left(t'<t_0\right)=0$]. We will also assume that the
observation time $t$ is past the end of the laser pulse.  The initial
wavefunction $\ket{\Psi^{(0)}}$ is a vibronic wavefunction, including both
electronic and nuclear degrees of freedom. In the last line, we replaced
$\hat{U}_{i}\left(t,t';E_{I}\right)$ by $\hat{U}_{IR}\left(t,t';E_{I}\right)$,
thus neglecting the possibility of absorbing additional XUV photons.

So far, we have avoided choosing a specific representation of the vibronic
wavefunctions. For the initial wavefunction $\Psi^{(0)}$, we use the standard
adiabatic Born-Huang Ansatz:
\begin{align}
  \ket{\Psi^{(0)}\left(t\right)} &=
     \sum_a \ket{\psi_a\left(r;q\right)} \ket{\chi_a\left(q,t\right)}, & \label{eqn:psi0}
\end{align}
where $\psi_a$ are the discrete, time-independent electronic states of the
neutral species, which depend on the electronic coordinates $r$ and
parametrically on the nuclear coordinates $q$. Time-dependent nuclear
wavepackets $\chi_a$ propagate on these electronic surfaces. We take
that the electronic states $\psi_a$ and the corresponding surfaces are
available to us through some other means. (If desired, e.g. for treating
the situation where the initial, neutral wavepacket finds itself in a
vicinity of a conical intersection, Eq.~\ref{eqn:psi0} can be taken as 
a diabatic vibronic Ansatz, with minimal changes to the treatment.)

We assume that the set of electronic states $\psi_a$ is complete with respect
to the action of the propagator $\hat{U}_{IR}$. Under this assumption, one can
define an identity operator $\hat{1}_N$, which can be inserted between
operators $\hat{\mu}_i$ and $\hat{U}_{IR}\left(t',t_0;E_N\right)$ in
Eq.~\eqref{eqn:psi1t}:
\begin{align}
  \hat{1}_N &= \sum_a \ket{\psi_a\left(r;q\right)}\bra{\psi_a\left(r;q\right)}, & \label{eqn:oneN}
\end{align}
where the brackets are understood to imply integration over electronic
coordinates $r$ alone. The result is:
\begin{widetext}
\begin{align}
  \ket{\Psi_{i}^{(1)}\left(t\right)} 
     &=
        -\frac{\ii}{2} \E^{-\ii\Phi_i +\ii E_N t_0}
         \sum_a 
         \int_{t_0}^{t} dt' 
         \E^{-\ii E_I t}
         \E^{\ii \left(E_{I}-E_{N}-\Omega_{i}\right) t'}
         f_{i}\left(t'\right)
         \hat{U}_{IR}\left(t,t';E_{I}\right)
         \hat{\mu}_{i} 
         \ket{\psi_a\left(r;q\right)}
         \ket{\chi_a\left(q,t'\right)}
         , & \label{eqn:psi1ta} \\
  \ket{\chi_a\left(q,t'\right)} &=
         \sum_b
         \hat{u}_{ab}\left(t',t_0;E_N\right)
         \ket{\chi_b\left(q,t_0\right)}
         , & \label{eqn:chia-t} \\
  \hat{u}_{ab}\left(t',t_0;E_N\right) &=
         \braket{\psi_a\left(r;q\right)|
            \hat{U}_{IR}\left(t',t_0;E_{N}\right)
            |\psi_b\left(r;q\right)}
         , & \label{eqn:hatu-ab}
\end{align}
\end{widetext}
where we have chosen to introduce vibrational propagator
$\hat{u}_{ab}\left(t',t_0;E_N\right)$.

The quantity $\ket{\chi_a\left(q,t'\right)}$ is to be understood as a
vibrational wavepacket on an electronic surface $a$ at time $t'$. We assume
that efficient means of propagating these wavepackets are available to us. In
the most common special case, where $\ket{\Psi^{(0)}\left(t_0\right)}$ is an
eigenstate of the field-free Hamiltonian with energy $E_N$, and the effects of
the IR field on the initial neutral wavefunction can be neglected,
Eq.~\ref{eqn:chia-t} reduces simply to:
\begin{align}
   \ket{\chi_a\left(q,t'\right)} &\overset{\textrm{G.S.}}{=} \ket{\chi_a\left(q,t_0\right)}
   . & \tag{\ref{eqn:chia-t}a}.
\end{align}

Under our assumptions, absorption of an XUV photon brings the molecule into a
highly-excited electronic state, with one of the electrons either ionized or in
a Rydberg state. If we assume that at most one electron is ionized or excited,
while the others remain tightly bound, it is natural to expand the wavefunction
after XUV absorption in the form:
\begin{align}
  \ket{\Psi^{(1)}\left(t\right)} &=
     \sum_c
     \int d k
     \ket{\psi_{ck}\left(r;q\right)}
     \ket{\chi_c\left(q,t\right)}
     , & \label{eqn:psi1}
\end{align}
where discrete index $c$ is understood to run over the
asymptotically-populated, tightly-bound states of the residual ion, while the
general index $k$ labels the full electronic state (continuum or discrete)
associated to this ion core. As before, functions $\psi_{ck}$ and the
corresponding energy surfaces are assumed to be available to us. 

Analogously to Eq.~\eqref{eqn:oneN}, we introduce identity-resolution operator
$\hat{1}_C$ in the ion space:
\begin{align}
  \hat{1}_C &= \sum_c \int d k \ket{\psi_{ck}\left(r;q\right)}\bra{\psi_{ck}\left(r;q\right)}. & \label{eqn:oneC}
\end{align}
Inserting $\hat{1}_C$ into Eq.~\eqref{eqn:psi1ta} between $\hat{U}_{IR}$ and
$\hat{\mu}_i$ and rearranging the terms, we obtain:
\begin{widetext}
\begin{align}
  \ket{\Psi_{i}^{(1)}\left(t\right)} &=
        -\frac{\ii}{2} \E^{-\ii\Phi_i +\ii E_N t_0}
         \sum_{a,c}
         \int d k 
         \int_{t_0}^{t} dt' 
         \E^{-\ii E_I t}
         \E^{\ii \left(E_{I}-E_{N}-\Omega_{i}\right) t'}
         f_{i}\left(t'\right)
         \hat{U}_{IR}\left(t,t';E_{I}\right)
         \ket{\psi_{ck}\left(r;q\right)}
         \hat{\mu}_{ck,a}
         \ket{\chi_a\left(q,t'\right)}
         , & \label{eqn:psi1tb}
\end{align}
\end{widetext}
\begin{align}
  \hat{\mu}_{ck,a}\left(q\right) &=
         \bra{\psi_{ck}\left(r;q\right)}
         \hat{\mu}_{i} 
         \ket{\psi_a\left(r;q\right)}
        , & \label{eqn:mucka}
\end{align}
where operator $\hat{\mu}_{ck,a}$ is the transition dipole for ionization (or
excitation) of an electronic state $\ket{\psi_a\left(r;q\right)}$, forming
state $\ket{\psi_{ck}\left(r;q\right)}$. This operator depends parametrically
on the nuclear coordinates $q$.

We can now introduce yet another identity-resolution operator $\hat{1}_{C'}$,
in the form:
\begin{align}
  \hat{1}_{C'} &= \sum_d \int d p \ket{\psi_{dp}\left(r;q\right)}\bra{\psi_{dp}\left(r;q\right)}, & \label{eqn:oneCprime}
\end{align}
where the (discrete) index $d$ and general parameter $p$ are understood as the
labels of the final state of the photoion and photoelectron, respectively.
Inserting $\hat{1}_{C'}$ to the left of the operator $\hat{U}_{IR}$ in
Eq.~\eqref{eqn:psi1tb} and rearranging, we obtain:
\begin{widetext}
\begin{align}
  \ket{\Psi_{i}^{(1)}\left(t\right)} &=
        -\frac{\ii}{2} \E^{-\ii\Phi_i +\ii E_N t_0}
         \sum_d 
         \int d p \;
         \E^{-\ii E_I t}
         \ket{\psi_{dp}\left(r;q\right)}
         \ket{\chi_{dpi}\left(q,t\right)}
         , & \label{eqn:psi1tc} \\
  \ket{\chi_{dpi}\left(q,t\right)} &=
         \sum_{a,c}
         \int d k 
         \int_{t_0}^{t} dt' 
         \E^{\ii \left(E_{I}-E_{N}-\Omega_{i}\right) t'}
         f_{i}\left(t'\right)
         \hat{u}_{dp,ck}\left(t,t';E_I\right)
         \hat{\mu}_{ck,a}
         \ket{\chi_a\left(q,t'\right)}
         , & \label{eqn:chidpi} \\
  \hat{u}_{dp,ck}\left(t,t';E_I\right) &=
         \bra{\psi_{dp}\left(r;q\right)}
         \hat{U}_{IR}\left(t,t';E_{I}\right)
         \ket{\psi_{ck}\left(r;q\right)}
        . & \label{eqn:hatu-dp-ck}
\end{align}
\end{widetext}
In Eq.~\eqref{eqn:chidpi}, $\chi_{dpi}\left(q,t\right)$ is the final amplitude
of the ion state $d$ and photoelectron state $p$, at nuclear coordinates $q$
and time $t$, generated by the XUV field $F_{i}$. Propagator
\eqref{eqn:hatu-dp-ck} describes evolution of the initially-prepared ionized
(or excited) state under the influence of the IR field. We note that the
meaning of the phase factor $\E^{-\ii E_{I}t}$ is subtly different between
Eqs.~\eqref{eqn:psi1tb} and \eqref{eqn:psi1tc}. In Eq.~\eqref{eqn:psi1tb}, it
is a global overall phase, while in Eq.~\eqref{eqn:psi1tc} $E_{I}$ is permitted
to be $p$-dependent. This change amounts to a gauge transformation of
$\ket{\chi_{dpi}\left(q,t\right)}$, which is compensated by the counteracting
transformation of the $\hat{u}_{dp,ck}$ propagator in Eq.~\eqref{eqn:chidpi}.

We would now like to examine the propagator
$\hat{u}_{dp,ck}\left(t,t';E_I\right)$ of Eq.~\eqref{eqn:hatu-dp-ck} a bit more
closely. Without any formal justification, we will now introduce the crucial
approximation of our treatment. We will \textit{assume} that:
\begin{widetext}
\begin{align}
  \hat{u}_{dp,ck}\left(t,t';E_I\right) &\approx
         \hat{u}_{p,k}\left(t,t';E_{I}-E_{C}\right)
         \hat{u}_{d,c}\left(t,t';E_{C}\right)
         , & \label{eqn:hatu-dp-ck-prod} \\
  \hat{u}_{d,c}\left(t,t';E_{C}\right) &=
         \bra{\psi_{d}\left(r;q\right)}
         \hat{U}_{IR}\left(t,t';E_{C}\right)
         \ket{\psi_{c}\left(r;q\right)}
         , & \label{eqn:hatu-dc} \\
  \hat{u}_{p,k}\left(t,t';E_{I}-E_{C}\right) &=
         \bra{\psi_{p}\left(r;q\right)}
         \hat{U}_{IR}\left(t,t';E_{I}-E_{C}\right)
         \ket{\psi_{k}\left(r;q\right)}
         , & \label{eqn:hatu-pk} \\
  \left[\hat{u}_{d,c},\hat{u}_{p,k}\right] &= 0
         . & \label{eqn:hatu-commutator}
\end{align}
\end{widetext}
In Eq.~\eqref{eqn:hatu-dc}, $E_{C}$ is the characteristic energy of the
cationic manifold, while $\ket{\psi_{c}\left(r;q\right)}$ are Born-Oppenheimer
electronic wavefunctions of the cation.  Similar to Eq.~\eqref{eqn:hatu-ab}, we
assume that efficient means of evaluating Eq.~\eqref{eqn:hatu-dc} are available
to us. Propagator \eqref{eqn:hatu-pk}, could have been formally (and
tautologically) defined as:
\begin{align}
  \hat{u}_{p,k}\left(t,t';E_{I}-E_{C}\right) &\overset{?}{=}
     \hat{u}_{dp,ck}\left(t,t';E_I\right) \hat{u}_{c,d}\left(t',t;E_{C}\right)
  . & \label{eqn:hatu-pk-formal}
\end{align}
(Note however that the putative definition \eqref{eqn:hatu-pk-formal} does not
satisfy the commutator relation \eqref{eqn:hatu-commutator}, except for the
trivial case of a single-state cationic manifold.)

From Eq.~\eqref{eqn:hatu-pk-formal}, it is clear that in adopting
Eq.~\eqref{eqn:hatu-dp-ck-prod}, we neglect the possibility of a transition in
the $(c,d)$ manifold inducing a transition in the $(k,p)$ manifold and vice
versa. Examples of such transitions are collisionally-induced transitions in
the ion core, as well as shake-off and shake-up transitions. By our initial
assumptions, the two manifolds (the compact ion core and the extended
photoelectron/Rydberg orbital) are energetically and spatially separated, so
that such transitions are expected to have low relative cross-sections.  If
necessary, they could be treated as higher-order perturbations.

We should also emphasize that partitioning of the Hamiltonian implied by the
Eq.~\eqref{eqn:hatu-dp-ck-prod} does \textit{not} introduce the single-particle
approximation. This can be clearly seen in the special case where the cationic
manifold consists of an isolated, non-degenerate ground state. Then, the
propagator $\hat{u}_{d,c}$ amounts to a $q$-dependent phase change, and
Eq.~\eqref{eqn:hatu-dp-ck-prod} is exact, rather than an approximation.
Indices $k,p$ then enumerate all electronic states in the system -- both
excited and ionized. The energies of these states must however be taken
relative to the energy of the target state, $E_{C}$.

We can now evaluate the propagator of Eq.~\eqref{eqn:hatu-pk}, treating the
IR field as a perturbation to the zeroth-order Hamiltonian:
\begin{widetext}
\begin{align}
   \hat{U}_{IR}\left(t,t';E_{k}\right) \ket{\psi_{k}\left(r;q\right)} &=
    \xcancel{
      \hat{U}_{0}\left(t,t';E_{k}\right) \ket{\psi_{k}\left(r;q\right)}
      }
    -\ii \int_{t'}^{t} d t" 
     \hat{U}_{IR}\left(t,t";E_{k}\right) 
     \hat{V}_{IR} 
     \hat{U}_{0}\left(t",t';E_{k}\right)
     \ket{\psi_{k}\left(r;q\right)}
   , & \label{eqn:IRk}
\end{align}
\end{widetext}
where $E_{k}=E_{I}-E_{C}$ is the ``excess'' characteristic energy of the state
$\ket{\psi_{kc}}$ relative to $E_{C}$ -- the characteristic energy of the
cation. The first term on the right-hand side of Eq.~\eqref{eqn:IRk} preserves
$k$, leading to the $\delta_{p-k}$ contribution upon substitution into
Eq.~\eqref{eqn:hatu-pk}.  Because we are interested in the RABBITT sidebands,
rather than in the main harmonic line, we can ignore this contribution from now
on. Then, substituting $\hat{V}_{IR}$ from Eq.~\eqref{eqn:VIR}, we obtain,
separately for emission ($+$) and absorption ($-$) or an IR photon:
\begin{widetext}
\begin{align}
  \hat{u}_{p,k,\pm}\left(t,t';E_{k}\right) 
    &\overset{\textrm{s.b.}}{=}
    -\frac{\ii}{2} 
     \int_{t'}^{t} d t" 
     f_{IR}\left(t"\right) 
     \E^{\pm\ii\omega t"} 
     \bra{\psi_{p}\left(r;q\right)}
     \hat{U}_{IR}\left(t,t";E_{k}\right) 
     \hat{\mu}_{IR}  
     \hat{U}_{0}\left(t",t';E_{k}\right)
     \ket{\psi_{k}\left(r;q\right)}
     & \nonumber \\
    & \approx
    -\frac{\ii}{2} 
     \int_{t'}^{t} d t" 
     f_{IR}\left(t"\right) 
     \E^{\pm\ii\omega t"} 
     \bra{\psi_{p}\left(r;q\right)}
     \hat{U}_{0}\left(t,t";E_{k}\right) 
     \hat{\mu}_{IR}  
     \hat{U}_{0}\left(t",t';E_{k}\right)
     \ket{\psi_{k}\left(r;q\right)}
     & \nonumber \\
    & =
    -\frac{\ii}{2} 
     \int_{t'}^{t} d t" 
     f_{IR}\left(t"\right) 
     \E^{\pm\ii\omega t"} 
     \E^{\ii\left(E_{k}-E_{p}\right) \left(t-t"\right)}
     \bra{\psi_{p}\left(r;q\right)}
     \hat{U}_{0}\left(t,t";E_{p}\right) 
     \hat{\mu}_{IR}  
     \hat{U}_{0}\left(t",t';E_{k}\right)
     \ket{\psi_{k}\left(r;q\right)}
     & \nonumber \\
    & =
    -\frac{\ii}{2} 
     \E^{-\ii\left(E_{p}-E_{k}\right) t}
     \int_{t'}^{t} d t" 
     f_{IR}\left(t"\right) 
     \E^{\ii\left(E_{p}-E_{k}\pm\omega\right) t"}
     \bra{\psi_{p}\left(r;q\right)}
     \hat{\mu}_{IR}  
     \ket{\psi_{k}\left(r;q\right)}
     & \nonumber \\
    & =
    -\frac{\ii}{2} 
     \E^{-\ii\left(E_{p}-E_{k}\right) t}
     \bra{\psi_{p}\left(r;q\right)}
     \hat{\mu}_{IR}  
     \ket{\psi_{k}\left(r;q\right)}
     \int_{t'}^{t} d t" 
     f_{IR}\left(t"\right) 
     \E^{\ii\left(E_{p}-E_{k}\pm\omega\right) t"}
    , & \label{eqn:hatu-pk-pm-raw}
\end{align}
\end{widetext}
where in the second line, we have neglected the possibility of absorbing the
second IR photon.  In the third line, we shifted the energy origin of the
left-most propagator and rearranged the terms.  In the fourth line, we have
used the assumption that $\ket{\psi_{k}}$ are eigenfunctions of the field-free
Hamiltonian with energy $E_{k}$, so that:
\begin{align}
  \hat{U}_{0}\left(t_2,t_1;E_{k}\right) \ket{\psi_{k}\left(r;q\right)} &= 
     \ket{\psi_{k}\left(r;q\right)}
  , & \label{eqn:U0psik}
\end{align}
and moved the IR-dipole matrix element outside of the integral.

To progress further, we now evaluate the $dt"$ integral in
Eq.~\eqref{eqn:hatu-pk-pm-raw} by parts, using adiabatic turn-on procedure
(see \S42 of Ref.~\cite{LLIII}):
\begin{widetext}
\begin{align}
    \int_{t'}^{t} d t" 
    f_{IR}\left(t"\right) 
    \E^{\lambda t"}
    \E^{\ii\left(E_{p}-E_{k}\pm\omega\right) t"}
    & =
    \int_{t'}^{t} d t" 
    f_{IR}\left(t"\right) 
    \frac{d}{d t"}
    \left(
      -\ii
      \frac{\E^{\lambda t" + \ii\left(E_{p}-E_{k}\pm\omega\right) t"}}
           {\left(E_{p}-E_{k}\pm\omega\right) -\ii \lambda}
    \right)
    & \nonumber \\
    & =
    -\ii
    f_{IR}\left(t"\right) 
      \frac{\E^{\lambda t" + \ii\left(E_{p}-E_{k}\pm\omega\right) t"}}
           {\left(E_{p}-E_{k}\pm\omega\right) -\ii \lambda}
      \Bigg|_{t'}^{\bcancel{t}}
    +\ii
    \int_{t'}^{t} d t" 
    \xcancel{\frac{d f_{IR}\left(t"\right)}{d t"}}
      \frac{\E^{\lambda t" + \ii\left(E_{p}-E_{k}\pm\omega\right) t"}}
           {\left(E_{p}-E_{k}\pm\omega\right) -\ii \lambda}
    & \nonumber \\
    & \approx
    \ii
    f_{IR}\left(t'\right) 
      \frac{\E^{\lambda t' + \ii\left(E_{p}-E_{k}\pm\omega\right) t'}}
           {\left(E_{p}-E_{k}\pm\omega\right) -\ii \lambda}
    & \nonumber \\
    & \overset{\lambda\rightarrow+0}{=}
    \ii
    f_{IR}\left(t'\right) 
      \frac{\E^{\ii\left(E_{p}-E_{k}\pm\omega\right) t'}}
           {\left(E_{p}-E_{k}\pm\omega\right) -\ii 0^+}
    , \label{eqn:fIR-int}
\end{align}
\end{widetext}
where we have used the slowly-varying envelope approximation to neglect the
time derivative of the envelope $f_{IR}$, and assumed that the observation time
$t$ is past the end of the IR pulse.

Inserting Eq.~\eqref{eqn:fIR-int} into Eq.~\eqref{eqn:hatu-pk-pm-raw}, we then
obtain our final expression for $\hat{u}_{p,k,\pm}$:
\begin{widetext}
\begin{align}
  \hat{u}_{p,k,\pm}\left(t,t';E_{k}\right) 
  & =
    \frac{1}{2} 
    \E^{-\ii\left(E_{p}-E_{k}\right) t}
    f_{IR}\left(t'\right) 
    \bra{\psi_{p}\left(r;q\right)}
    \hat{\mu}_{IR}  
    \ket{\psi_{k}\left(r;q\right)}
    \frac{\E^{\ii\left(E_{p}-E_{k}\pm\omega\right) t'}}
         {\left(E_{p}-E_{k}\pm\omega\right) -\ii 0^+}
    . & \label{eqn:hatu-pk-pm}
\end{align}
\end{widetext}

Substituting Eqs.~\eqref{eqn:hatu-pk-pm} and \eqref{eqn:hatu-dp-ck-prod} into
Eqs.~(\ref{eqn:psi1tc}--\ref{eqn:chidpi}), we then obtain for the second-order
wavefunction response $\Psi_{i\pm}^{(2)}$, where we have chosen to separate 
contributions due to emission and absorption of the IR photons:
\begin{widetext}
\begin{align}
  \ket{\Psi_{i\pm}^{(2)}\left(t\right)} &=
        -\frac{\ii}{2} \E^{-\ii\Phi_i +\ii E_N t_0}
         \sum_d 
         \int d p \;
         \E^{-\ii \left(E_{C}+E_{p}\right) t}
         \ket{\psi_{dp}\left(r;q\right)}
         \E^{+\ii \left(E_{p}-E_{k}\right) t}
         \ket{\chi_{dpi\pm}\left(q,t\right)}
         , & \nonumber \\
       &=
        -\frac{\ii}{2} 
         \E^{-\ii\Phi_i -\ii E_{C} t +\ii E_N t_0}
         \sum_{d}
         \int d p \;
         \E^{-\ii E_{p} t}
         \ket{\psi_{dp}\left(r;q\right)}
         \ket{\tilde{\chi}_{dpi\pm}\left(q,t\right)}
         , & \label{eqn:psi2t} \\
  \ket{\tilde{\chi}_{dpi\pm}\left(q,t\right)} &=
         \E^{+\ii \left(E_{p}-E_{k}\right) t}
         \sum_{a,c}
         \int d k 
         \int_{t_0}^{t} dt' 
         \E^{\ii \left(E_{C}+E_{k}-E_{N}-\Omega_{i}\right) t'}
         f_{i}\left(t'\right)
         \frac{1}{2} 
         \E^{-\ii\left(E_{p}-E_{k}\right) t}
         f_{IR}\left(t'\right) 
         \bra{\psi_{p}\left(r;q\right)}
         \hat{\mu}_{IR}  
         \ket{\psi_{k}\left(r;q\right)}
       & \nonumber \\
       & \times
         \frac{\E^{\ii\left(E_{p}-E_{k}\pm\omega\right) t'}}
         {\left(E_{p}-E_{k}\pm\omega\right) -\ii 0^+}
         \hat{u}_{d,c}\left(t,t';E_{C}\right)
         \hat{\mu}_{ck,a}
         \ket{\chi_a\left(q,t'\right)}
         & \nonumber \\
       & =
         \frac{1}{2} 
         \sum_{a,c}
         \int_{t_0}^{t} dt' 
         \E^{\ii \left(E_{p}+E_{C}-E_{N}-\Omega_{i}\pm\omega\right) t'}
         f_{i}\left(t'\right)
         f_{IR}\left(t'\right) 
         \hat{u}_{d,c}\left(t,t';E_{C}\right)
         \int d k 
         \frac{
         \bra{\psi_{p}\left(r;q\right)}
         \hat{\mu}_{IR}  
         \ket{\psi_{k}\left(r;q\right)}
         \hat{\mu}_{ck,a}
         }
         {\left(E_{p}-E_{k}\pm\omega\right) -\ii 0^+}
         \ket{\chi_a\left(q,t'\right)}
         & \nonumber \\
       & =
         \frac{1}{2}
         \sum_{a,c}
         \int_{t_0}^{t} dt' 
         \E^{-\ii \epsilon_{cp,a\pm} t'}
         f_{i}\left(t'\right)
         f_{IR}\left(t'\right) 
         \hat{u}_{d,c}\left(t,t';E_{C}\right)
         \hat{D}_{cp,a\pm}\left(q\right)
         \ket{\chi_a\left(q,t'\right)}
         , & \label{eqn:tildechi-dp} \\
  \epsilon_{cp,a\pm} &= 
         \left(\Omega_{i}\mp\omega\right)
       - \left(E_{p}+E_{C}-E_{N}\right)
         , & \label{eqn:epsilon-cpa} \\
  \hat{D}_{cp,a\pm}\left(q\right) &=
         \int d k 
         \frac{\hat{\mu}_{p,ck} \hat{\mu}_{ck,a}}
              {\left(E_{p}-E_{k}\pm\omega\right) -\ii 0^+}
         , & \label{eqn:D-cpa} \\
  \hat{\mu}_{p,ck}\left(q\right) &= 
         \bra{\psi_{p}\left(r;q\right)}
         \hat{\mu}_{IR}  
         \ket{\psi_{k}\left(r;q\right)}
         , & \label{eqn:mu-pck}
\end{align}
\end{widetext}
where we used $E_{I}=E_{k}+E_{C}$, and moved the phase contribution in
$\ket{\chi_{dpi}}$ dependent solely on $t$ into the definition of
$\Psi_{i}^{(2)}$. 

The individual terms in Eqs.~(\ref{eqn:psi2t}--\ref{eqn:mu-pck}) have a
transparent physical interpretation.  The quantity $\left(E_{p}+E_{C}\right)$
is the total electronic energy of the final state of the system.  The
non-trivial dynamics in the system is described by nuclear wavepacket(s)
$\ket{\tilde{\chi}_{dpi\pm}}$, which propagate on ionic surface $d$ and are
entangled with final photoelectron momentum $p$. The operator
$\hat{D}_{cp,a\pm}$ is the standard electronic matrix element for 2-photon
absorption. The quantity $\epsilon_{cp,a\pm}$ is the amount of energy deposited
into the nuclear degrees of freedom of the system. Finally,
Eq.~\eqref{eqn:tildechi-dp} describes time evolution of the nuclear wavepacket
on the (generally coupled) ionic energy surfaces. The Fourier transform in
Eq.~\eqref{eqn:tildechi-dp} picks out the relevant spectral component of the
nuclear wavepacket.

\subsection{RABBITT signal: General case}

Using Eqs.~(\ref{eqn:psi2t}--\ref{eqn:tildechi-dp}), we are ready to describe
the RABBITT sidebands. Sideband $M$ arises due to interference between
two-photon transitions involving two neighboring harmonics:
\begin{align}
  \Omega_1 & = (M-1)\omega, & \label{eqn:Omega1} \\
  \Omega_2 & = (M+1)\omega, & \label{eqn:Omega2}
\end{align}
The signal at final photoelectron momentum $p$ is given by a sum of four
contributions:
\begin{align}
  I\left(p\right) &= 
    I_{1-,1-}\left(p\right)
    + I_{2+,2+}\left(p\right)
    + I_{1-,2+}\left(p\right)
    + I_{2+,1-}\left(p\right)
  . & \label{eqn:rabbitt}
\end{align}
The first contribution is the photoelectron signal due to the simultaneous
absorption of an $\Omega_1$ and $\omega$ photons.  The second term describes
absorption of an $\Omega_2$ and emission of an $\omega$ photons. The two
remaining terms, which form a complex-conjugate pair, is the delay-dependent
interference term.

From Eq.~\eqref{eqn:psi2t}, the individual contributions are given by:
\begin{widetext}
\begin{align}
  I_{l,r}\left(p\right) 
     &=
       \braket{\Psi_{l}^{(2)}\left(t\right)|p'}\braket{p|\Psi_{r}^{(2)}\left(t\right)}
       \Big|_{p'\rightarrow p}
     & \nonumber \\
     &=
        \left(
        -\frac{\ii}{2} 
         \E^{-\ii\Phi_{l} -\ii E_{C} t +\ii E_N t_0}
         \sum_{c}
         \E^{-\ii E_{p'} t}
         \ket{\psi_{cp'}\left(r;q\right)}
         \ket{\tilde{\chi}_{cp'l}\left(q,t\right)}
        \right)^\dagger & \nonumber \\
        & \times
        \left(
        -\frac{\ii}{2} 
         \E^{-\ii\Phi_{r} -\ii E_{C} t +\ii E_N t_0}
         \sum_{d}
         \E^{-\ii E_{p} t}
         \ket{\psi_{dp}\left(r;q\right)}
         \ket{\tilde{\chi}_{dpr}\left(q,t\right)}
        \right)
       \Big|_{p'\rightarrow p}
       & \nonumber \\
     &=
        \frac{1}{4} 
        \E^{\ii \left(\Phi_l - \Phi_r\right)}
        \sum_{cd}
         \bra{\tilde{\chi}_{cp'l}\left(q,t\right)}
         \bra{\psi_{cp'}\left(r;q\right)}
           \E^{\ii\left(E_{p'}-E_{p}\right)t}
         \ket{\psi_{dp}\left(r;q\right)}
         \ket{\tilde{\chi}_{dpr}\left(q,t\right)}
       \Big|_{p'\rightarrow p}
       & \nonumber \\
     &=
        \frac{1}{4} 
        \E^{\ii \left(\Phi_l - \Phi_r\right)}
        \sum_{cd}
         \bra{\tilde{\chi}_{cp'l}\left(q,t\right)}
         \delta_{cd}
         \delta\left(\frac{\vec{p'}-\vec{p}}{2\pi}\right)
         \ket{\tilde{\chi}_{dpr}\left(q,t\right)}
       \Big|_{p'\rightarrow p}
       & \nonumber \\
     &=
        \frac{1}{4} 
        \delta\left(\frac{\vec{p'}-\vec{p}}{2\pi}\right)
        \E^{\ii \left(\Phi_l - \Phi_r\right)}
        \sum_{d}
         \braket{\tilde{\chi}_{dpl}\left(q,t\right)|
                 \tilde{\chi}_{dpr}\left(q,t\right)}
       , & \nonumber 
\end{align}
\end{widetext}
where $l,r=1-,2+$, and we assumed that the continuum functions are normalized
to $\delta\left(\frac{\vec{p'}-\vec{p}}{2\pi}\right)$. Any other normalization
choice will lead to an equivalent expression, provided that a consistent choice
is make in Eq.~\eqref{eqn:D-cpa}. We will therefore omit the continuum
normalization factor from now on.  Substituting $\ket{\tilde{\chi}_{dpi\pm}}$
from eq.~\eqref{eqn:tildechi-dp}, we get:
\begin{widetext}
\begin{align}
  I_{l,r}\left(p\right) 
     &=
        \frac{1}{4} 
        \E^{\ii \left(\Phi_l - \Phi_r\right)}
        \sum_{d}
        \left(
           \frac{1}{2}
           \sum_{b,e}
           \int_{t_0}^{t} dt" 
           \E^{-\ii \epsilon_{ep,bl} t"}
           f_{l}\left(t"\right)
           f_{IR}\left(t"\right) 
           \hat{u}_{d,e}\left(t,t";E_{C}\right)
           \hat{D}_{ep,bl}\left(q\right)
           \ket{\chi_b\left(q,t"\right)}
        \right)^\dagger
        & \nonumber \\
        & \times
        \left(
           \frac{1}{2}
           \sum_{a,c}
           \int_{t_0}^{t} dt' 
           \E^{-\ii \epsilon_{cp,ar} t'}
           f_{r}\left(t'\right)
           f_{IR}\left(t'\right) 
           \hat{u}_{d,c}\left(t,t';E_{C}\right)
           \hat{D}_{cp,ar}\left(q\right)
           \ket{\chi_a\left(q,t'\right)}
        \right)
     & \nonumber \\
     &=
        \frac{1}{16} 
        \E^{\ii \left(\Phi_l - \Phi_r\right)}
        \sum_{b,e,a,c}
        \int_{t_0}^{t} dt" 
        \int_{t_0}^{t} dt' 
        \E^{+\ii \epsilon_{p} \left(t"-t'\right)}
        f_{l}\left(t"\right)
        f_{IR}\left(t"\right) 
        f_{r}\left(t'\right)
        f_{IR}\left(t'\right) 
        & \nonumber \\
        & \times 
        \bra{\chi_b\left(q,t"\right)}
        \hat{D}_{ep,bl}^\dagger\left(q\right)
        \left(
          \sum_{d}
          \hat{u}_{e,d}\left(t",t;E_{C}\right)
          \hat{u}_{d,c}\left(t,t';E_{C}\right)
        \right)
        \hat{D}_{cp,ar}\left(q\right)
        \ket{\chi_a\left(q,t'\right)}
     & \label{eqn:Ilrp-intermediate}
\end{align}
\end{widetext}
where we used $\epsilon_{ep,bl}=\epsilon_{cp,ar}=\epsilon_p$, which holds due
to our choice of $\Omega_1$ and $\Omega_2$, and $\hat{u}_{d,e}
\left(t,t";E_{C}\right)^\dagger = \hat{u}_{e,d} \left(t",t;E_{C}\right)$. We
further use the ion-state completeness assumption to replace:
\begin{widetext}
\begin{align}
  \sum_{d}
  \hat{u}_{e,d}\left(t",t;E_{C}\right)
  \hat{u}_{d,c}\left(t,t';E_{C}\right)
  & =
  \hat{u}_{e,c}\left(t",t';E_{C}\right)
  & \label{eqn:ion-completeness}
\end{align}
giving:
\begin{align}
  I_{l,r}\left(p\right) 
     &=
        \E^{\ii \left(\Phi_l - \Phi_r\right)}
        {\tilde I}_{l,r}\left(p\right) 
      ,& \label{eqn:Ilrp}
\end{align}
\begin{align}
  {\tilde I}_{l,r}\left(p\right) 
     &=
        \frac{1}{16} 
        \int_{t_0}^{t} dt" 
        \int_{t_0}^{t} dt' 
        \E^{+\ii \epsilon_{p} \left(t"-t'\right)}
        f_{l}\left(t"\right)
        f_{IR}\left(t"\right) 
        f_{r}\left(t'\right)
        f_{IR}\left(t'\right) 
        & \nonumber \\
        & \times
        \sum_{a,b,c,e}
        \bra{\chi_b\left(q,t"\right)}
        \hat{D}_{ep,bl}^\dagger\left(q\right)
        \hat{u}_{e,c}\left(t",t';E_{C}\right)
        \hat{D}_{cp,ar}\left(q\right)
        \ket{\chi_a\left(q,t'\right)}
      .& \label{eqn:tildeIlrp}
\end{align}
\end{widetext}
The quantity ${\tilde I}_{l,r}$ is the ``intrinsic'' part of the RABBITT matrix
element, which does not depend on the relative delay of the XUV and the IR
fields. The entire delay dependence is encapsulated by the phase pre-factor
$\E^{\ii \left(\Phi_l - \Phi_r\right)}$ in Eq.~\eqref{eqn:Ilrp}.  As expected
on physical grounds, the initial and observation times ($t_0$ and $t$) drop out
of the final expression, provided that the envelopes of all pulses are zero
outside of the $[t_0:t]$ interval. These integration limits can therefore be
replaced by $[-\infty:+\infty]$ if desired.  Equations \eqref{eqn:rabbitt},
\eqref{eqn:Ilrp}, and \eqref{eqn:tildeIlrp} are our general-case result for 
the RABBITT spectrum. Below, we consider some of the relevant special cases.

\section{RABBITT signal: Special cases\label{sec:special-cases}}

Although the result of Eq.~\eqref{eqn:tildeIlrp} is compact and physically
transparent, it is invokes a rather complex object: a weighted two-time nuclear
cross-correlation function [$\braket{\chi_b|...|\chi_a}$]. We would like to
consider possible simplifications to Eq.~\eqref{eqn:tildeIlrp}. For 
weak IR fields and short pulses, it is reasonable to neglect 
vibrational excitation by the IR field can be neglected, both in the
neutral and in the cationic manifolds. Then, the propagator $\hat{u}_{e,c}$
can be replaced by the field-free propagator $\hat{u}^{0}_{e,c}$, which is
invariant with respect to shift of the time origin:
\begin{align}
  \hat{u}_{e,c}\left(t",t';E_{C}\right) 
         &\approx
         \hat{u}^{0}_{e,c}\left(t"-t';E_{C}\right) & \label{eqn:hatu-ec-approx} \\
  \hat{u}^{0}_{e,c}\left(\tau;E_{C}\right) &=
         \bra{\psi_{e}\left(r;q\right)}
         \hat{U}_{0}\left(\tau,0;E_{C}\right)
         \ket{\psi_{c}\left(r;q\right)}
         . & \label{eqn:hatu0-ec}
\end{align}
Furthermore, the initial, neutral vibronic wavefunction in many stable molecules
is well-represented by a single-surface Born-Oppenheimer product, so that the
$a,b$ sums in Eq.~\eqref{eqn:tildeIlrp} collapse to a single, time-independent 
term:
\begin{align}
  \ket{\chi_a\left(q,t\right)}
        &\approx
        \ket{\chi_0\left(q\right)}
        .& \label{eqn:chi0}
\end{align}

Additionally, the characteristic decay time scale for the cationic
autocorrelation functions is often small, on the order of a few femtoseconds
or tens of femtoseconds\cite{Tannor2006,Patchkovskii2009}. On these time
scales, the difference between pulse envelopes at $t'$ and $t"$ can be
neglected (CW approximation). Furthermore, in the diabatic representation, the
cationic cross-correlation functions remain small on the timescale of the IR
and XUV pulse duration, and only the diabatic autocorrelations need to be
considered:
\begin{align}
  \hat{u}^{0}_{e,c}\left(\tau;E_{C}\right) 
      &\approx
      \delta_{ec} 
      \hat{u}^{0}_{c,c}\left(\tau;E_{C}\right) 
      \equiv
      \hat{u}^{0}_{c}\left(\tau;E_{C}\right) 
      .& \label{eqn:hatu0-ec-dia}
\end{align}

Applying the approximations above to the general Eq.~\eqref{eqn:Ilrp}, we
obtain:
\begin{widetext}
\begin{align}
  I_{l,r}\left(p\right) 
     &=
        \E^{\ii \left(\Phi_l - \Phi_r\right)}
        P_{l,r}
        \sum_{c}
           M_{c,l,r}\left(p\right)
      , & \label{eqn:IlrSimple} \\
   P_{l,r} &=
        \frac{\pi}{8} 
        \int d \tau
           f_{l}\left(\tau\right)
           f_{r}\left(\tau\right)
           f_{IR}^2\left(\tau\right) 
     , & \label{eqn:Plr} \\
   M_{c,l,r}\left(p\right) 
     &=
        \frac{1}{2\pi}
        \int d\tau
        \E^{+\ii \epsilon_{p} \tau}
           \bra{\chi_0\left(q\right)}
           \hat{D}_{cp,0l}^\dagger\left(q\right)
           \hat{u}^{0}_{c}\left(\tau;E_{C}\right)
           \hat{D}_{cp,0r}\left(q\right)
           \ket{\chi_0\left(q\right)}
      , & \label{eqn:Mclr}
\end{align}
\end{widetext}
where time integrals are over all times, and we have chosen to apply
normalization factor $\frac{1}{2\pi}$ to the definition of the matrix element
$M_c\left(p\right)$, for reasons which will become clear in the following
section.  In Eq.~\eqref{eqn:IlrSimple}, the pulse-envelope parameters
($P_{l,r}$) are cleanly separated from the molecular factors ($M_{c,l,r}$),
while the entire time-delay dependence is encapsulated by the phase prefactor
$\E^{\ii\left(\Phi_l-\Phi_r\right)}$, similar to the familiar atomic case\cite{Paul2001}.
While at the first glance, the conditions under which the approximate
Eq.~\eqref{eqn:IlrSimple} is obtained do appear very restrictive, a closer
examination shows that they are expected to be satisfied for a large fraction
of small, rigid molecules.

It is instructive to examine some limiting cases of Eq.~\eqref{eqn:Mclr}.

\subsection{No nuclear motion: The ``atomic'' case}

We first consider the case of nuclear motion being entirely absent in the centre-of-mass coordinate system,
so that the molecule has well-defined coordinates $q_{0}$, which are unchanged by the action of the
propagator $\hat{u}^{0}_{c}$. Then:
\begin{align}
   M_{c,l,r}\left(p\right) 
     &\overset{\textrm{fixed}}{=}
        \hat{D}_{cp,0l}^\dagger\left(q_{0}\right)
        \hat{D}_{cp,0r}\left(q_{0}\right)
        \frac{1}{2\pi}
        \int d\tau
        \E^{+\ii \epsilon_{p} \tau}
     & \nonumber \\
     &=
        \hat{D}_{cp,0l}^\dagger\left(q_{0}\right)
        \hat{D}_{cp,0r}\left(q_{0}\right)
        \delta\left(\epsilon_{p}\right)
      . & \label{eqn:Mclr-fixed}
\end{align}
This result coincides with the familiar atomic case: The RABBITT sidebands
appear at the photoelectron energy of $M\omega-IP$, where $IP$ is the
ionization potential. Their width is determined by the combined bandwidth of
the XUV and IR pulses. Obviously, no isotope dependence is possible in this
approximation.

\subsection{Nuclear motion: Condon approximation}

The next natural approximation to consider is to treat the electronic part of
the matrix element \eqref{eqn:Mclr} as $q$-independent: the Condon
approximation. Then:
\begin{align}
   M_{c,l,r}\left(p\right) 
     &\overset{\textrm{Condon}}{=}
        \hat{D}_{cp,0l}^\dagger\left(q_{0}\right)
        \hat{D}_{cp,0r}\left(q_{0}\right)
        N_{c}\left(\epsilon_{p}\right)
      , & \label{eqn:Mclr-condon} \\
   N_{c}\left(\epsilon_{p}\right)
     & =
        \frac{1}{2\pi}
        \int d\tau
        \E^{+\ii \epsilon_{p} \tau}
           A_{c}\left(\tau;E_{C}\right)
      , & \label{eqn:N-c} \\
   A_{c}\left(\tau;E_{C}\right)
     & =
           \bra{\chi_0\left(q\right)}
           \hat{u}^{0}_{c}\left(\tau;E_{C}\right)
           \ket{\chi_0\left(q\right)}
      , & \label{eqn:A-c}
\end{align}
where $N_{c}$ is the Fourier transform of nuclear auto-correlation function on
(diabatic) cationic surface $c$. The field-free autocorrelation function is
Hermitian with respect to time reversal\cite{Tannor2006}:
\begin{align}
    A_{c}\left(-\tau\right) &= A_{c}\left(\tau\right)^\dagger
   , & \label{eqn:AC-hermiticity}
\end{align}
so that $N_{c}$ is guaranteed to be real. As long as the time integration
domain in Eq.~\eqref{eqn:N-c} is not truncated, $N_{c}$ is also guaranteed to
be positive semi-definite\cite{Tannor2006}:
\begin{align}
   N_{c}\left(\epsilon_{p}\right) \ge& \;0. & \label{eqn:N-c-positivity}
\end{align}

An important consequence of Eq.~\eqref{eqn:N-c-positivity} is that vibrational
dynamics in the Condon approximation \textit{does not introduce additional time
delays} in the RABBITT spectrum. Time delays remain an exclusively electronic
property in this approximation, similar to the atomic case. Nuclear motion
however imposes a finite, intrinsic photoelectron energy profile onto the
RABBIT spectrum.  In the Condon approximation, this profile coincides with the
vibration profile in 1-photon photoionization spectrum at the same
photoelectron energy.

Isotopic dependence can arise in two ways in this approximation. First, the
initial wavepacket $\ket{\chi_0}$ and the propagator $\hat{u}^{0}_{c}$ depend
on the nuclear masses, so that the factor $N_{c}$ is isotope-dependent. This
contribution affects the photoelectron-energy profile, but not the time delays
or the contrast of the signal oscillations with time. Additional isotope
dependence could arise if the characteristic geometries $q_{0}$, where the
electronic matrix elements are determined, are not the same for the 
isotopomers involved.

\subsection{Nuclear motion: Zero-point effects}

In the Condon approximation [Eq.~\eqref{eqn:Mclr-condon}], the electronic part
of the matrix element $M_{c,l,r}$ is evaluated at a single, characteristic
geometry $q_0$. A natural refinement is to consider the consequences of the
finite spatial extent of the wavepacket, by averaging the electronic matrix
element over the initial wavepacket. All nuclear wavepackets will have non-zero
spatial extent due to the effects of the zero-point motion. Vibrational
excitation will also affect the extent of the wavepacket. If the overall shape
of the wavepacket, apart from the central position, is unaffected by nuclear
motion, we obtain:
\begin{align}
   M_{c,l,r}\left(p\right) 
     &\overset{\textrm{Z.P.E.}}{=}
        G_{c,l,r}\left(p\right)
        N_{c}\left(\epsilon_{p}\right)
      , & \label{eqn:Mclr-zpe} \\
   G_{c,l,r}\left(p\right)
     &=
        \bra{\chi_0\left(q\right)}
          \hat{D}_{cp,0l}^\dagger\left(q\right)
          \hat{D}_{cp,0r}\left(q\right)
        \ket{\chi_0\left(q\right)}
      , & \label{eqn:Gclrp}
\end{align}
where $N_{c}\left(\epsilon_{p}\right)$ is given by Eq.~\eqref{eqn:N-c} above.
For the most important special case, where $\ket{\chi_0}$ is the ground-state
vibrational wavefunction of a multi-dimensional harmonic oscillator, the
integral \eqref{eqn:Gclrp} can be readily evaluated. 

Indeed, for a 1-dimensional harmonic oscillator of unit effective mass and
force constant $\omega_i^2$:
\begin{align}
  \left(
     -\frac{1}{2}\frac{\partial^2}{\partial q_i^2}
     +\frac{1}{2}\omega_i^2 q_i^2 - \frac{1}{2}\omega_i
  \right) \chi_{0,i}\left(q_i\right) & = 0
  , & \label{eqn:harmonic}
\end{align}
\begin{align}
  \chi_{0,i}\left(q_i\right) &= 
    \left(\frac{\omega_i}{\pi}\right)^{\frac{1}{4}} \E^{-\frac{\omega_i}{2}q_i^2}
  , & \label{eqn:harmonic-zpfi} \\
  \ket{\chi_0\left(q\right)}
    &= \prod_{i} \chi_{0,i}\left(q_i\right)
  , & \label{eqn:harmonic-zpf}
\end{align}
where $q_i$ is the displacement from the equilibrium position $q_{0,i}$ and
$\omega_i$ is the vibrational quantum. The multidimensional vibrational ground
state is a product of $\chi_{0,i}$ for all modes. The first few non-zero
moments of $\chi_{0,i}$, which are required below, are given by:
\begin{align}
  \int d q_i \chi_{0,i}^2\left(q_i\right)     &= 1,                       \label{eqn:chi0i-q0} \\
  \int d q_i q_i^2 \chi_{0,i}^2\left(q_i\right) &= \frac{1}{2\omega_i},   \label{eqn:chi0i-q2} \\
  \int d q_i q_i^4 \chi_{0,i}^2\left(q_i\right) &= \frac{3}{4\omega_i^2}. \label{eqn:chi0i-q4}
\end{align}
The classical turning points of the ground-state vibrational wavefunction 
of mode $i$ are found at $q_i = \pm \omega_i^{-\frac{1}{2}}$.

As long as matrix elements $\hat{D}$ are sufficiently smooth, they can be
expanded in Taylor series:
\begin{align}
  \hat{D}_{cp,0x}
     &\approx
        D^{(0)}_x
      + \sum_{i} D^{(i)}_x q_i
      + \frac{1}{2} \sum_{ij} D^{(i,j)}_x q_i q_j
     , & \label{eqn:Dcpx-taylor} \\
  D^{(0)}_x   & = \hat{D}_{cp,0x}\left(q_0\right), & \label{eqn:D0x} \\
  D^{(i)}_x   & = \frac{\partial}{\partial q_i} 
                  \hat{D}_{cp,0x}\left(q_0\right), & \label{eqn:D0x-i} \\
  D^{(i,j)}_x & = \frac{\partial^2}{\partial q_i \partial q_j} 
                  \hat{D}_{cp,0x}\left(q_0\right), & \label{eqn:D0x-ij} 
\end{align}

Inserting Eqs.~\eqref{eqn:harmonic-zpf} and \eqref{eqn:Dcpx-taylor} in
Eq.~\eqref{eqn:Gclrp}, we obtain:
\begin{widetext}
\begin{align}
   G_{c,l,r}\left(p\right)
     &\overset{Z.P.E.}{=}
       \iiint_{-\infty}^{+\infty}
       \prod_{k} 
          d q_k
       \prod_{i,j} 
          \left(\frac{\omega_i}{\pi}\right)^{\frac{1}{4}} 
          \E^{-\frac{\omega_i}{2}q_i^2}
          \left(\frac{\omega_j}{\pi}\right)^{\frac{1}{4}} 
          \E^{-\frac{\omega_j}{2}q_j^2}
       & \nonumber \\
       & \times
       \left(
           D^{(0)}_l
         + \sum_{k} D^{(k)}_l q_k
         + \frac{1}{2} \sum_{ko} D^{(k,o)}_l q_k q_o
       \right)^\dagger
       \left(
           D^{(0)}_r
         + \sum_{m} D^{(m)}_r q_m
         + \frac{1}{2} \sum_{mn} D^{(m,n)}_r q_m q_n
       \right)
       & \nonumber \\
     &=
       \iiint_{-\infty}^{+\infty}
       \prod_{k} 
          d q_k
       \prod_{i,j} 
          \left(\frac{\omega_i}{\pi}\right)^{\frac{1}{4}} 
          \E^{-\frac{\omega_i}{2}q_i^2}
          \left(\frac{\omega_j}{\pi}\right)^{\frac{1}{4}} 
          \E^{-\frac{\omega_j}{2}q_j^2}
       & \nonumber \\
       & \times
       \Bigg[
           \left(D^{(0)}_l\right)^\dagger D^{(0)}_r
         + \left(D^{(0)}_l\right)^\dagger \frac{1}{2} \sum_k D_r^{(k,k)} q_k^2
         + \sum_k \left(D^{(k)}_l\right)^\dagger D^{(k)}_r q_k^2
         + \frac{1}{2}\sum_{k} \left(D^{(k,k)}_l\right)^\dagger D^{(0)}_r q_k^2
       & \nonumber \\
       & + \frac{1}{4}\sum_{k} \left(D^{(k,k)}_l\right)^\dagger D^{(k,k)}_r q_k^4
         + \frac{1}{4}\sum_{k\ne m} \left(D^{(k,k)}_l\right)^\dagger D^{(m,m)}_r q_k^2 q_m^2
         + \frac{1}{2}\sum_{k\ne m} \left(D^{(k,m)}_l\right)^\dagger D^{(k,m)}_r q_k^2 q_m^2
       \Bigg]
       & \nonumber \\
     & =
           \left(D^{(0)}_l\right)^\dagger D^{(0)}_r
       & \nonumber \\
       & + 
           \sum_{k} \frac{1}{4\omega_k} \left[
                  \left(D^{(0)}_l\right)^\dagger D_r^{(k,k)}
              + 2 \left(D^{(k)}_l\right)^\dagger D^{(k)}_r
              +   \left(D^{(k,k)}_l\right)^\dagger D^{(0)}_r
              \right]
       & \nonumber \\
       & + \sum_{k} \frac{3}{16\omega_k^2} \left(D^{(k,k)}_l\right)^\dagger D^{(k,k)}_r
         + \sum_{k\ne m} \frac{1}{16\omega_k\omega_m} 
                  \left(D^{(k,k)}_l\right)^\dagger D^{(m,m)}_r
         + \xcancel{
           \sum_{k\ne m} \frac{1}{8\omega_k\omega_m} 
                  \left(D^{(k,m)}_l\right)^\dagger D^{(k,m)}_r
           }    
    , & \label{eqn:GclrpZPE}
\end{align}
\end{widetext}
where we used parity arguments to drop vanishing contributions containing odd
powers of any of the $q_i$ coordinates. The last three terms in
Eq.~\eqref{eqn:GclrpZPE} are of the $4$-th order in $q$, higher than the formal
order of Eq.~\eqref{eqn:Dcpx-taylor}. The two terms containing the diagonal
part of the second-derivative matrix of $D$ ensure that the approximated
$G_{c,l,r}$ remains positive semidefinite. These terms must be kept to obtain
physically meaningful results. The last contribution to
Eq.~\eqref{eqn:GclrpZPE}, which is expensive to evaluate, can be safely
omitted.

If analytical derivatives of the matrix elements in
Eq.~\eqref{eqn:GclrpZPE} are not available, they can be obtained using the
standard finite-difference formulae. It is particularly convenient to use 
the turning points of the normal modes. Then:
\begin{align}
  D^{(k)}_x &\approx \frac{\sqrt{\omega_k}}{2} 
             \left( D_x\left( \sqrt{\omega_k}\right)
                  - D_x\left(-\sqrt{\omega_k}\right) 
             \right)
     , & \label{eqn:Dkx-finite} \\
  D^{(k,k)}_x &\approx \omega_k
             \left(   D_x\left( \sqrt{\omega_k}\right)
                  +   D_x\left(-\sqrt{\omega_k}\right) 
                  - 2 D_x\left( q_0            \right) 
             \right)
     , & \label{eqn:Dk2x-finite}
\end{align}
where $x=l,r$. Substituting into Eq.~\eqref{eqn:GclrpZPE}, we then obtain the
final working expression:
\begin{widetext}
\begin{align}
   G_{c,l,r}\left(p\right)
     &\overset{Z.P.E. FD}{=}
                           D_l    ^\dagger D_r
         +   \sum_k        D_l    ^\dagger W_{r,k}
         +   \sum_k        W_{l,k}^\dagger D_r
         +   \sum_k        V_{l,k}^\dagger V_{r,k}
         + 3 \sum_k        W_{l,k}^\dagger W_{r,k}
         +   \sum_{k\ne m} W_{l,k}^\dagger W_{r,m}
     , & \label{eqn:GclrpFD} \\
   D_x &= \hat{D}_{cp,0x}\left(q_c\right)
     , & \label{eqn:ZPE-Dx} \\
   V_{x,k} 
     & = 
       \frac{1}{\sqrt{8}}
              \left[
                 \hat{D}_{cp,0x}^\dagger\left(q_0+\sqrt{\omega_k}\right)
               - \hat{D}_{cp,0x}^\dagger\left(q_0-\sqrt{\omega_k}\right)
              \right]
     , & \label{eqn:ZPE-Vxk} \\
   W_{x,k}
     & =
       \frac{1}{4}
           \left[
                 \hat{D}_{cp,0x}\left(q_0+\sqrt{\omega_k}\right)
              +  \hat{D}_{cp,0x}\left(q_0-\sqrt{\omega_k}\right)
              - 2\hat{D}_{cp,0x}\left(q_0\right)
              \right]
     . & \label{eqn:ZPE-Wxk}
\end{align}
\end{widetext}
Eq.~\eqref{eqn:GclrpFD} can be applied to the situation where the nuclear
motion after ionization is negligible. In this case, the characteristic
geometry $q_c$ and the neutral equilibrium geometry $q_0$ coincide. If the
characteristic geometry does not coincide with the neutral equilibrium
geometry, Eq.~\eqref{eqn:GclrpFD} still guarantees that the matrix element
$G_{c,l,r}$ remains positive semidefinite.

\section{Summary and outlook\label{sec:summary}}

In this work, we develop the formal theory of RABBITT photoionization spectra
in molecular systems. Our most general result is given by
Eqs.~\eqref{eqn:rabbitt}, \eqref{eqn:Ilrp}, \eqref{eqn:tildeIlrp},
\eqref{eqn:D-cpa}, and \eqref{eqn:hatu-dc}. It includes the effects of the
nuclear motion in the initial (neutral) and final (cation) state (including
vibrational heating by the IR field and IR-induced electronic transitions);
coordinate dependence of the two-electron photoionization matrix elements; and
effects of the finite pulse duration.  It neglects the possibility of
collisional excitation of the final ion, as well as of the shake-off and
shake-up processes. The possibility of multiphoton transitions due to the XUV,
or absorption/emission of multiple IR photons by the ionized electron are also
neglected. The vibronic dynamics, including dynamics at conical intersections
is treated fully, both in the initial and the final molecular states.

We further analyze important special cases. We demonstrate that in the
lowest-order, Condon approximation, nuclear motion does not introduce
additional time delays in the RABBITT spectra. The photoelectron energy profile
in this approximation coincides with the vibrational profile in 1-photon
ionization spectra. In contrast, the zero-point motion leads to non-vanishing
phase contributions, and therefore time delays.  We develop numerically
tractable expressions for the ZPE contributions, both for the case where
analytical derivatives of the electronic matrix elements are available, and for
the finite-difference evaluation.

The expressions we have developed can be readily evaluated by combining the
existing molecular photoionization codes, utilizing fixed-nuclei
approximation, \textit{ab initio} potential energy surfaces, and molecular
vibronic-dynamics simulations. We envision routine applications of our theory
to molecules with tens and potentially hundreds vibrational degrees of freedom,
which are entirely out of reach for brute-force, coupled electron-nuclear
simulations. Work in this direction is currently underway, and will be reported
elsewhere.

\bibliography{rabbitt-molecules}

\end{document}